\documentclass[aps, notitlepage, nofootinbib, twocolumn]{revtex4-1}

\usepackage{geometry}
\usepackage{epsfig}
\usepackage{epstopdf}
\usepackage{bm}
\usepackage[nooneline]{caption}

\begin{document}

\title{Reality, No Matter How You Slice It}
\author{Ken Wharton}
\affiliation{Department of Physics and Astronomy, San Jos\'{e} State University, San Jos\'{e}, CA 95192-0106}
\begin{abstract}

\setlength{\baselineskip}{1.1\baselineskip} 

In order to reject the notion that information is always \textit{about something}, the ``It from Bit'' idea relies on the nonexistence of a realistic framework that might underly quantum theory. This essay develops the case that there \textit{is} a plausible underlying reality: one actual spacetime-based history, although with behavior that appears strange when analyzed dynamically (one time-slice at a time).  By using a simple model with \textit{no} dynamical laws, it becomes evident that this behavior is actually quite natural when analyzed ``all-at-once'' (as in classical statistical mechanics). The ``It from Bit" argument against a spacetime-based reality must then somehow defend the importance of dynamical laws, even as it denies a reality on which such fundamental laws could operate.

\end{abstract}

\maketitle

\setlength{\baselineskip}{1.25\baselineskip} 

\captionsetup[figure]{justification=raggedright}

\section{Introduction}

Information, not so long ago, used to always mean knowledge \textit{about something}.   Even today, under layers of abstraction, that's still the usual meaning.\footnote{The technical concept of Shannon Information is distinct from this everyday meaning, although they are often erroneously conflated.  Shannon Information is perhaps better termed ``source compressibility'' or ``channel capacity'' (in different contexts), and is a property of (real) sources or channels. \cite{Timpson}  This essay utilizes the everyday meaning of ``information": an agent's knowledge.}  Sure, an agent can be informed of a string of bits (via some signal) without knowing what the bits refer to, but at minimum the agent has been informed about the physical signal itself.

Quantum theory, however, has led many to question this once-obvious connection between knowlege/information and an underlying reality.  Not only is our information about a quantum system indistinguishable from our best physical description, but we have failed to come up with a realistic account of what might be going on \textit{independent} of our knowledge.  This blurring between information and reality has led to a confusion as to which is more fundamental.

The remarkable ``It from Bit'' idea \cite{Wheeler} that \textit{information} is more fundamental than reality is motivated by standard quantum theory, but this is a bit suspicious.  After all, there's a long ``instrumentalist'' tradition of only using what we can measure to describe quantum entities, rejecting outright any story of what might be happening when we're not looking.  Using a theory that only comprises our knowledge of measurement outcomes to justify knowledge as fundamental is almost like wearing rose-tinted glasses to justify that the world is tinted red.

But any such argument quickly runs into the counterargument: ``Then answer the question: What \textit{is} the (objective) reality that our information of quantum systems is actually \textit{about}?"  Without an answer to this question (that differs from our original information), ``It from Bit" proponents can perhaps claim to win the argument by default.  The only proper rebuttal is to demonstrate that there is some plausible underlying reality, after all.

This is generally thought to be an impossible task, having been ruled out by various ``no-go'' theorems \cite{Bell,KS,PBR}.  But such theorems are only as solid as their premises, and they all presume a particular sort of independence between the past and the future.  This presumption may be valid in a universe that uses dynamical laws to evolve some initial state into future states, but if ``The Universe is Not a Computer'' \cite{FQXi4}, there is a natural alternative to this dynamic viewpoint.  As argued in last year's essay, instead of the universe solving itself one time-slice at a time, it's possible that it only looks coherent when solved ``all-at-once''.

This essay aims to demonstrate how this all-at-once perspective naturally recasts our supposedly-complete information about quantum systems into \textit{incomplete} information about an underlying, spacetime-based reality.  After some motivation in the next section, a simple model will demonstrate how the all-at-once perspective works for purely spatial systems (without time).  Then, applying the same perspective to spacetime systems will reveal a framework that can plausibly serve as a realistic explanation for quantum phenomena.

The result of this analysis will be to dramatically weaken the ``It from Bit" idea, showing that it's possible to have an underlying reality, even in the case of quantum theory.  We may still choose to reject this option, but the mere fact that it is on the table might encourage us \textit{not} to redefine information as fundamental -- especially as it becomes clear just how poorly-informed we actually are.

\section{Instants vs. Spacetime}

The case for discarding dynamics in favor of an all-at-once analysis is best made by analyzing quantum theory \cite{FQXi4}, but it's also possible to frame this argument using the \textit{other} pillar of modern physics: Einstein's theory of relativity.  The relevant insight is that there is no objective way to slice up spacetime into instants, so we must not assign fundamental significance to any particular slice.

Figure 1 is a standard spacetime diagram (with one dimension of space suppressed).  If run forward in time like a movie, this diagram represents two spatial objects that begin at a common past (C.P.) and then move apart.  But if viewed all-at-once, the figure instead shows two orange ``worldtubes" that intersect in the past.  In relativity, as we are about to see, it is best to analyze this picture all-at-once.

The most counter-intuitive feature of special relativity is that there is no objective ``now''.  Simultaneous events for one observer are not simultaneous for another.  No observer is right or wrong; ``now'' is merely subjective, not an element of reality.  An illustration of this can be seen in Figure 1.  Observer \#1 has a ``now" that slices the worldtubes into two white ovals, while Observer \#2 has a ``now" that slices the worldtubes into two black ovals.  Clearly, they disagree.

\begin{figure}[htb]
\centerline{\includegraphics[width=.5\textwidth]{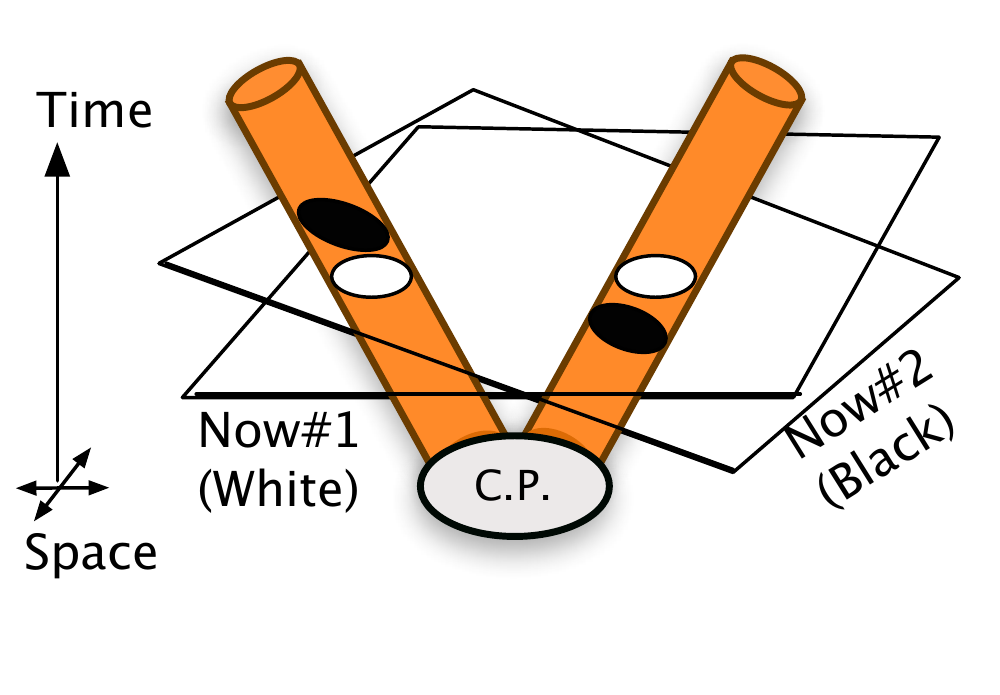}}
\caption{A spacetime diagram, demonstrating the unreality of ``now''.  (See text.)}
\label{Figure:Fig1}
\end{figure}

This fact implies that any dynamical movie made from a spacetime diagram will incorporate a subjective choice of how to slice it up.  One way to purge this subjectivity is to simply view a spacetime diagram as a single 4D block.  After all, with no objective ``now'', there is no objective line between the past and the future, meaning there can be no objective difference between them.

Such a claim is counter-intuitive, \textit{but this is a central lesson of relativity.}  The only difference between the future and the past, in this view, is subjective: we don't (yet) know any of our future.  Arguments such as ``But the future isn't real \textit{now}" are no more meaningful than arguing ``Over there isn't real right here".

A more reasonable fallback for the dynamicist is not to deny that spacetime \textit{can} be viewed as a single 4D block, but rather to note that if dynamical equations govern the universe\footnote{Along with other subtleties, such as the existence of Cauchy data.} then \textit{any} complete spacelike slice suffices to generate the rest of the block (via dynamical equations).  So while no one slice is special, they're all equally valid inputs from which the full universe can be recovered.  Taken to an extreme, this viewpoint leads to the notion that the 4D block is filled with redundant permuted copies of the same 3D slice.  It also forbids a number of solutions allowed by general relativity, spacetime geometries warped to such an extent that they only make sense all-at-once.

The other problem with this sliced perspective is that it all but gives up on objectivity.  Even if it's \textit{possible} to generate the block from a single slice (a point I'll dispute later on), how can one 3D slice truly generate the others if it is a subjective choice?  In Figure 1, if \textit{both} the white ovals and the black ovals are different complete descriptions of the same reality, it's the 4D worldtubes they generate that makes them consistent.  The clearest objective reality requires a bigger picture.

This point becomes even clearer when one introduces (subjective) uncertainty.  Suppose each of the worldtubes in Figure 1 represent a (temporally extended) shoebox, each containing a single shoe.  Also suppose that you knew the shoes were a matched pair, but not which shoe (R or L) was in which box (1 or 2).  To represent your information about the two boxes after they had separated (say, the white ovals in Figure 1), you might use an equal-probability mix of both possibilities:  $S_{mix}=[50\%(L_1R_2), 50\%(L_2R_1)]$.  This is not a potential state of reality, but a state in a larger ``configuration space" that weights possibilities that \textit{do} fit in spacetime.  Note that having less knowledge forces a more complicated description, even if the underlying reality is assuredly either $L_1R_2$ or $L_2R_1$.  

For these restricted-knowledge situations, the all-at-once viewpoint is invaluable if we are to make sense of what is going on when we open a shoebox and learn which shoe is inside.  If we take the dynamic view that we need only keep track of the 3D white ovals in Figure 1 to describe the entire 4D system, then $S_{mix}$ might seem to give us everything we need; from it we can compute outcome probabilities and the correlations between the two boxes.  Upon learning that (say) the left shoe is in box 1, we can even update our knowledge of the 3D state $S$ to the appropriate $[100\%(L_1R_2)]$.  But what is lost in this viewpoint is the mechanism for the updating; if our entire description is that of the 3D white ovals, this updating process might appear nonlocal, as if some spooky influence at box 1 is influencing the reality over at box 2.  

Sure, we know that nothing spooky is going on in the case of shoes, but that's only because we already know there's an underlying reality of which $S_{mix}$ represents (subjective) \textit{information}.  If the existence of an underlying reality is doubt (as in quantum theory), then analysis of the 3D state $S_{mix}$ cannot address whether anything spooky is happening.  To resolve that question, one has to look at the entire 4D structure.  All at once.

In the all-at-once viewpoint, after finding the left shoe in box 1 we update our local knowledge to $L_1$ (updating occurs when we learn new information).  But thinking in 4D, we also update our knowledge of the \textit{past}; we now know that that $L_1$ back in the C.P.  This in turn implies $R_2$ back in the C.P, and this allows us to update our knowledge of $R_2$ in the present.  It's the continuous link, via the past, that proves that we did not change the contents of box 2; it contained the right shoe all along.  Throw away the analysis of the 4D link, and there's no way to be sure.

Before moving on, it's worth noting that this classical story cannot explain all quantum correlations; in fact, it's exactly the story ruled out by a no-go theorem \cite{Bell}.  Such theorems generally start from the classical premise that we can assign subjective probabilities $p_i$ to possible 3D realities, $W_i$.  States of classical information then naturally take the form $S=[p_1(W_1), p_2(W_2),...,p_N(W_N)]$, a function on $3N$-dimensional configuration space.  (Note the probabilities are all subjective; only one particular $W$ is real; the rest are not.)  The quantum no-go theorems have proven that such a state cannot explain quantum measurements without some classically-impossible feature, such as negative probabilities.

The standard thinking is that since any workable version of $S$ cannot be classical information, it must be a new kind of reality in its own right.  Effectively, the standard view\footnote{Including both deBroglie-Bohm \cite{dBB} and Everettian \cite{Everett} approaches.} extends reality from spacetime to configuration space.  But an alternative option, explored below, is that reality merely requires an extension from 3D to 4D, along with an all-at-once analysis.  At this point it's probably not obvious how anything might change if the $W$'s spanned 4D spacetime, but that's because the standard dynamical viewpoint makes any such extension trivial.  (Thanks to dynamics, all the interesting information is always encoded in a 3D slice).  Exploring this option therefore requires jettisoning dynamics.

Still, old habits die hard; it's difficult to think about time without also thinking in terms of dynamical equations.  The 4D block is a good start, but it's time to demonstrate how it can be used to make physical predictions.  Fortunately, it's a standard procedure, so straightforward that it's nearly trivial.

\section{A Dynamics-Free Model}

Physicists know how to do physics without dynamics, because we can analyze 3D systems for which there are no dynamics, by definition.  A particularly useful approach is found in classical statistical mechanics, because in that case we never know the exact microscopic details, allowing us to deal with restricted knowledge situations.

The basic approach works like this.  First, determine the possible underlying realities; call each one a ``microstate'' $W_i$.  The key next step\footnote{Sometimes known as the ``fundamental postulate of statistical mechanics".} is to assign each $W_i$ an equal \textit{a priori} probability, $p_i$.  (Initially treat all possible states as equally likely.)  If we learn new information -- say, that $W_{9}$ is ruled out -- we set $p_{9}\!=\!0$ and renormalize the remaining probabilities such that they sum to 1.  Finally, we can determine the probability that the system has any particular feature by simply adding the probabilities of the microstates with that feature.  One \textit{could} introduce dynamics on top of this framework, but it's not a logical necessity.  

\begin{figure}[htb]
\centerline{\includegraphics[width=.5\textwidth]{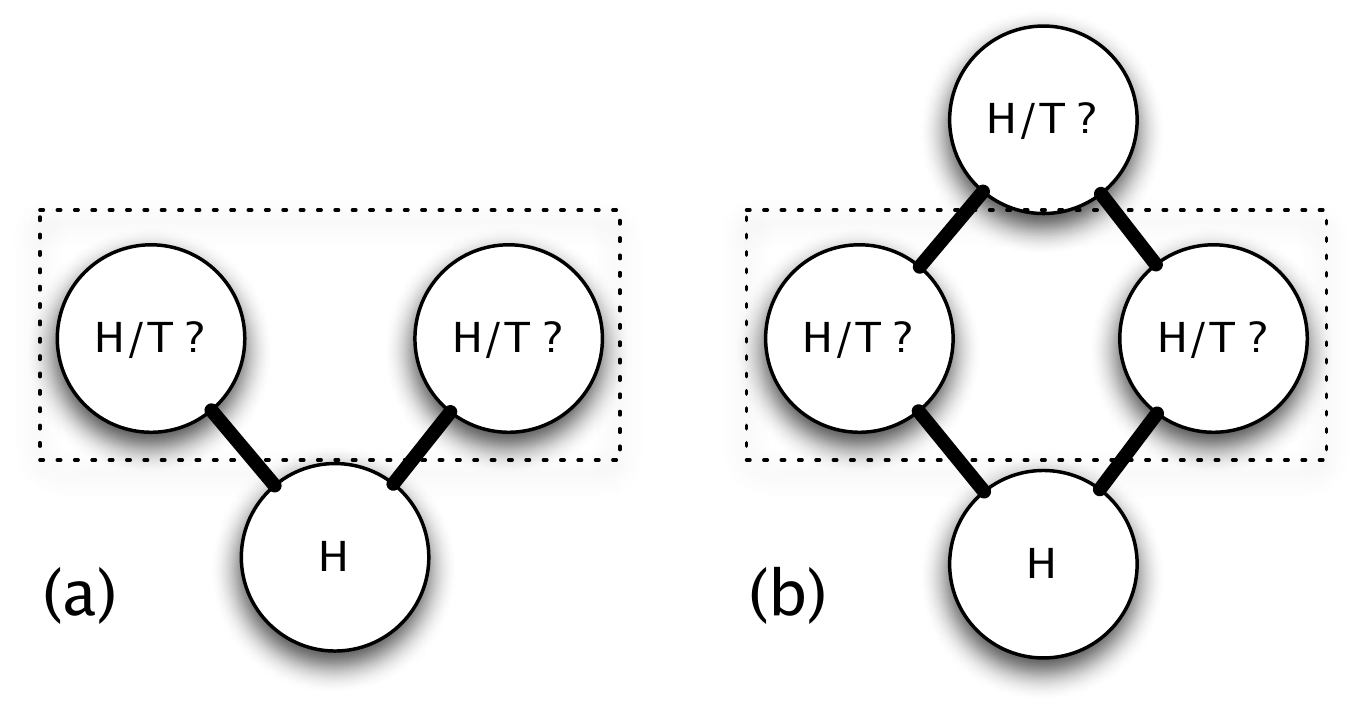}}
\caption{Two geometries of a model, in which each circle can be Heads (H) or Tails (T).  There are two line colors that connect matching circles (HH,TT) and a third line color to connect opposite circles (HT, TH).  The interesting case is where one does not know whether the geometry is that of 2a or 2b.}
\label{Figure:Fig2}
\end{figure}

For a simple example that will prove particularly relevant to quantum theory, consider Figure 2.  Each circle (perhaps a coin) can be in the state heads (H) or tails (T), and every line connects two circles.  Each line has one of three internal colors; red, green or blue, but these colors are \textit{unobservable} (they can sometimes be deduced, but not directly measured).  The model's only ``law" is that red lines must connect opposite-state circles ($H\!-\!T$ or $T\!-\!H$), while blue and green lines must connect similar-state circles  ($H\!-\!H$ or $T\!-\!T$).\footnote{This is effectively a much-simplified version of the Ising Model.}

Consider the following puzzle in the statistical mechanics framework:  In Figure 2a, if it is known that the bottom circle is $H$, what is the probability that the two circles in the dotted box are in the same state?  It's easy enough to work out (see the Appendix for details) that there are 4 microstates where these two circles are $HH$, 2 microstates for $HT$, 2 for $TH$, and only 1 for $TT$.  By assigning each of these 9 states an equal probability, it should be evident that there is a 5/9 chance those two circles are the same, and a 4/9 chance that they're different.

For Figure 2b, the same puzzle is trickier because now there's a fourth circle.  In this case, the same style of analysis (in the Appendix) reveals that the different geometry changes the probabilities.  In place of a 5:4 probability ratio, here one finds a 25:16 ratio.

The most interesting example is a further restriction where one \textit{does not know} whether the actual geometry is that of Figure 2a or 2b.  Specifically, one knows that the bottom circle is heads, and that the next two circles are connected, but not whether a fourth circle is connected (2b), or whether it is not (2a).

This is not to say there is no fact of the matter; there \textit{is} some particular geometry -- it's just unknown.  This is not quite the same as the unknown circles or links (which also have some particular state), because this model provides no clues as to how to calculate the probability of a \textit{geometry}.  All allowable states may be equally likely, but that doesn't help us if we don't know which states are allowable in the first place.  With this further-restricted knowledge, we would most naturally use an even higher-level configuration space to describe the probability of similar states in the dotted box, something like: $S_{?}\!\!=$[If 2a then 5/9; If 2b then 25/(25+16)].\footnote{There is nothing quite like this higher-level space in quantum theory, but as we'll see, that's the point.}

The next section will explore a crucial mistake that would lead one to conclude that \textit{no underlying reality exists} for this statistical-mechanics-based model, despite the fact that an underlying reality does indeed exist (by construction).  Then, by applying the above logic to a dynamics-free scenario in space \textit{and} time, we'll see how we are making this same mistake in quantum theory.  

\section{Implications of the Model}
\subsection{The Independence Fallacy}

Given the previous model, one might reasonably want to analyze a ``slice" of the system (the dotted box) independent from the rest.  But this can only be done by expanding the description of the box such that it includes \textit{each} possible external geometry -- effectively removing the ``ifs" from $S_{?}$.  Then, one can later use the actual geometry to extract the appropriate probability from the larger state space (either 5/9 or 25/41).

But this new perspective becomes quite mistaken if one further demands that the state of reality in the dotted box must \textit{be} independent of the external geometry.  It's obviously not true for this model, but given such an ``Independence Fallacy" one would be led to some interesting conclusions.  Namely, this all-possible-geometry state space would seem to be irreducible to a classical probability distribution over realistic microstates. 

Given the Independence Fallacy, the argument would go like this:  Geometry 2a implies a 5/9 probability of similar circles, while geometry 2b implies a 25/41 probability.  But since the state must be independent of the geometry, the question ``Are the two circles the same?" cannot be assigned a coherent probability.  And if it cannot be answered, such a question should not even be asked.

This, of course, is nonsense: such a question \textit{can} be asked in this model, but the answer depends on the geometry.  It is the Independence Fallacy which leads to a denial of an underlying reality -- stemming from a motivation to describe a slice of a system independently from what lies outside. 

\subsection{Information-Based Updating}

Leaving aside the Independence Fallacy, it should be clear how the $S_{?}$ description of the dotted box should be updated upon learning new information.  For example, if an agent learned that the geometry was in fact that of Figure 2b, a properly-updated description would simply be a 25/41 probability that the two coins were the same.  And upon learning the actual values of the coins (say, $HT$), further updating would occur; $HT$ would then have a 100\% probability.  

But the central point is that some information-updating naturally occurs when one learns the geometry of the model, even without any revealed circles.  And because this is a realistic model (with some real, underlying state), the information updating has no corresponding feature in the coin's objective reality.  It is a subjective process, performed as some agent gains new information.

\subsection{Introducing Time}

The above model was presented as a static system in two spatial dimensions.  The only place that time entered the analysis was in the updating process in the previous subsection, but this subjective updating had no relation to anything objective about the system.  Indeed, one could give different agents information in a different logical order, leading to different updating.  Both orders would be unrelated to any objective evolution of the system; after all, the system is static.

Still, an objective time coordinate can be introduced in a trivial manner: simply redefine the model such that one of the spatial axes in Figure 2 represents time instead of space.  Specifically, suppose that the vertical axis is time (past on the bottom, future on the top).  It is crucial not to introduce dynamics along with time; one point of the model was to show how to analyze systems without dynamics.  And since this analysis has already been performed, we don't need to do it again.  The dotted box now represents an instantaneous slice, and the same state-counting logic will lead to exactly the same probabilities as the purely spatial case.

One might be tempted to propose reasons why this space-time model is fundamentally different from the original space-space model, perhaps assuming the existence of dynamical laws.  Such laws \textit{would} break the analogy, but they are not part of the model.  Besides, the previous section is an existence proof that such a system \textit{can} be analyzed in this manner, which is all that is needed for the below conclusions.  It is logically possible to assign an equal probability to each temporally-extended microstate (or more intuitively, ``microhistory'')  and then make associated predictions.  

Sure, it's an open question whether there is some \textit{other} way to analyze systems without dynamics, or if this approach has any chance of actually making \textit{good} predictions.  But this approach \textit{is} empirically successful for spatial systems without dynamics, and the early indications are that it looks promising for temporal systems as well \cite{Wharton13}.

One unusual feature of the original model should now be obvious.  Not knowing the spatial geometry (say, 2a or 2b) was an artificial restriction.  But it's quite natural not to know the future, and once the vertical axis represents time, it's obvious why an agent might be uncertain whether the fourth circle would ever materialize.  But this does not break the analogy between the spatial and temporal models.  Sure, we tend to learn about things in temporal order, but it's not a formal requirement; we can film a movie of a system and watch it backwards, or even have spatial slices of a system delivered to us one at a time.\footnote{As in the final section of \cite{EPW}.}  The link between information-order and temporal-order is merely typical, not a logical necessity.

In this temporal context, it's also more understandable how one might fall into the Independence Fallacy.  If we expect the future to be generated from the past via some dynamical laws, then we would also expect the probabilities we assign to the past to be independent of the future experimental geometry.  But without dynamics, if we assign every microhistory an equal probability, the standard information-updating that made sense in the spatial case also makes sense in the temporal case.  When we learn about the experimental geometry of the future, this all-at-once analysis typically updates our probabilistic assessment of the past.

\section{Quantum Reality}

The modern arguments against an underlying reality for quantum systems typically involve hard-to-summarize ``no-go theorems'', but the central issues do not require anything so difficult, and indeed were well known to the founders of quantum mechanics.  A useful framework is the famous Double Slit Experiment, which in Richard Feynman's words reveals the ``central mystery" of quantum theory. \cite{Feynman} 

In Figure 3, a source (at the bottom) creates a single photon that passes up through a pair of slits.\footnote{The vertical axis is performing double-duty as both time and a second spatial axis.} The classical concept most closely related to photons are classical electromagnetic waves/fields, but photons behave in a way that disagrees with the dynamical Maxwell equations which govern such fields.  (A strike against dynamics.)  Namely, photons always seem to be measured in particle-like chunks, rather than spread out as classically predicted.  For example, when a lens (or two) images the slits (as in Figure 3a), one always finds that the photon-wave went through one slit \textit{or} the other.  

\begin{figure}[htb]
\centerline{\includegraphics[width=.5\textwidth]{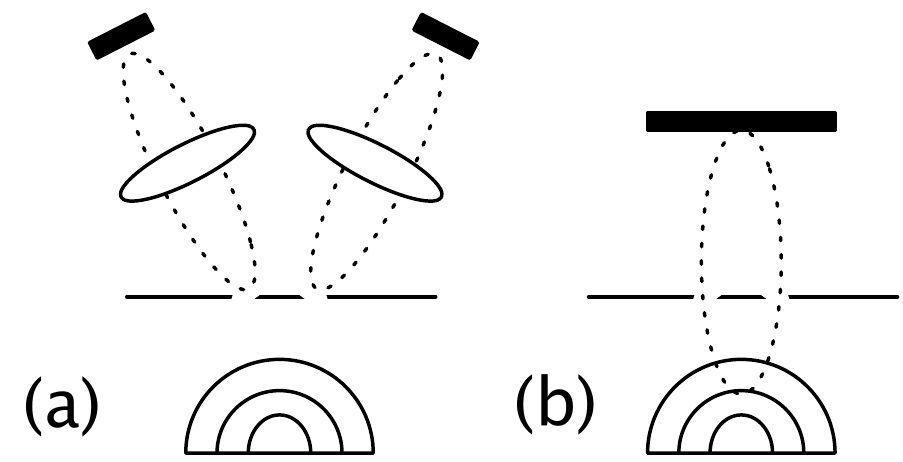}}
\caption{Two geometries of a double slit experiment, in which a single photon passes through a pair of slits.  3a) Lenses and (black) detectors measure which slit the photon passes through; 3b) A screen records a photon that contributes to a two-slit interference pattern.}
\label{Figure:Fig3}
\end{figure}

And yet it appears that photons \textit{do} spread out, at least between measurements, if one considers the experiment in Figure 3b.  Here a screen records the interference pattern produced by waves passing through \textit{both} slits, built up one photon at a time.  In the many-photon limit, this pattern is predicted by classical dynamics \textit{only} if the waves pass through both slits and interfere.  Since each individual photon conforms to this pattern (not landing in dark fringes), it seems evident that each photon also passes through both slits.

Where reality seems to fail here is the description of the photon at the slits -- one slice of the full spacetime diagram.  In 3a the photon seems to go through only one slit; in 3b it seems to go through both.  And since the status of the photon at the slits is ``obviously'' independent of the future experimental geometry, it follows that the actual location(s) of the photon-wave at the slits cannot be assigned a coherent probability.  

Except that this is \textit{exactly} the Independence Fallacy!  Compare Figure 2 (temporal version) to Figure 3; they are quite analogous.  In 2a and 3a the right and left branches stay separate; in 2b and 3b the geometry begins in the same way, but then allows recombination.  Following the above logic, \textit{avoiding} the Independence Fallacy allows a coherent underlying reality for the double-slit experiment.  

The answer is something like [If 3a then 50\%(left),50\%(right); If 3b then 100\%(both)].  Upon learning the future geometry, an agent would update her assessment of the past probabilities, just as before.  Once this updating occurs, a classical reality is revealed.  (For 3b, it is perfectly realistic to have a wave go through both slits.)  It only looks strange if you \textit{don't} analyze it all-at-once, or attempt to map this process onto a story with dynamical evolution.

Unlike other resolutions of the double-slit experiment, \textit{this} resolution naturally resolves more problematic situations.  The no-go theorems against realistic models all use the Independence Fallacy in one form or another.\footnote{Outcome independence \cite{Shimony}, preparation independence \cite{PBR}, etc.}  The typical assumption is that it's always fair to describe spatial slices independently from the future experimental geometry.  But if one updates past probabilities upon learning which measurement a system will encounter, the premises behind these theorems are explicitly violated.

Even so, this complicated updating of probabilities on different time-slices is not the most natural picture.  Relativity tells us that the slicing is subjective; the objective structure lies in the 4D spacetime block.  It is here where the microhistories reside, and to be realistic, one of these microhistories must \textit{really be there}.  A physics experiment is then about \textit{learning} which microhistory actually occurs, via information-based updating; we gain relevant information upon preparation, measurement setting, and measurement itself.  And the best way to coherently describe this updating is with an all-at-once analysis.

\section{Conclusions}

If there is a plausible reality underlying quantum theory, the ``It from Bit'' idea looks wrongheaded.  The microhistory-reality proposed here demands that one gives up the intuitive universe-as-computer story of dynamical time evolution, so one may still choose to cling to dynamics, voiding this analysis.  But in the process, one is also rejecting a spacetime-based reality.  Is this a fair trade-off?  Is dynamics really so crucial that it's worth delving into some nebulous ``informational immaterialism'' \cite{Timpson} or elevating configuration space into some weird reality in its own right?  And why should dynamical laws be so important if one is giving up on a fundamental reality in the first place?

After all, there are excellent reasons for dropping dynamics, the quantum no-go theorems being prime examples.  We also have the beautiful path integral where \textit{all} possible histories must be considered (whether they obey dynamical laws or not; see, \textit{e.g.}, \cite{WMP}).  And is it really so crucial that we live in a universe where nothing interesting happens in the time-direction, where everything about the present was encoded in some initial cosmic wavefunction?  It's not such a stretch to view our world as one \textit{possibility} of infinitely many, unshackled from strict predeterministic rules.

After giving up on reality via the Independence Fallacy, the standard quantum story ironically responds by making almost everything \textit{interdependent} in some strange configuration space.  (Almost everything, just not the future or the past.)  The simpler alternative proposed here is simply to link everything together in standard 4D spacetime.  This casts our information in the classical form: $S=[p_1(W_1), p_2(W_2),...,p_N(W_N)]$, with the crucial caveat that the $W$'s are now micro\textit{histories}, spanning 4D instead of 3D.  So long as one does not additionally impose dynamical laws, there is no theorem that one of these microhistories cannot be real. 

Still, qualitative arguments are one thing; the analogy between the above model and the double slit experiment can only be pushed so far.  And one can go \textit{too} far in the no-dynamics direction: considering \textit{all} histories, as in the path integral, would lead to the conclusion that the future would be almost completely uncorrelated with the past, contradicting macroscopic observations.  

But this approach can be made much more quantitative.  The key is to only consider a large natural subset of possible histories,\footnote{Those for which the total Lagrangian density is always zero.} such that classical dynamics is usually recovered as a general guideline in the many-particle limit.  Better yet, for at least one model, the structure of quantum probabilities naturally emerges.\footnote{The Born rule can be derived for measurements on an arbitrary spin state in reasonable limits \cite{Wharton13}.}  And as with any deeper-level theory that purports to explain higher-level behavior, intriguing new predictions are also indicated. \cite{Wharton13}

Even if the arguments presented in this essay are not a convincing reason to discard fundamental dynamical equations, they nevertheless serve as a strong rebuttal to the ``It from Bit'' proponents.  Whether or not one \textit{wants} to give up dynamics, the point is that one \textit{can} give up dynamics, in which case quantum information can plausibly be \textit{about something real}.  Instead of winning the argument by default, then, ``It from Bit'' proponents now need to argue that it's \textit{better} to give up reality.  Everyone else need simply embrace entities that fill ordinary spacetime -- no matter how you slice it.

\onecolumngrid

\clearpage

\twocolumngrid

 \section*{Appendix}
 
The model in Figure 2 (reproduced below) has the following rules.  Each circle can be in the state heads (H) or tails (T), and each line connects two circles.  Each line has one of three internal colors; red (R), green (G), or blue (B), but these colors are \textit{unobservable}.  The model's only ``law" is that red lines must connect opposite-state circles ($H\!-\!T$ or $T\!-\!H$), while blue and green lines must connect similar-state circles  ($H\!-\!H$ or $T\!-\!T$).

When analyzing the state-space, the key is to remember that connecting links between same-state circles have two possible internal colors (G or B), while links between opposite-state circles only have one possible color (R).  Combined with the equal \textit{a priori} probability of each complete microstate (both links and circles), this means that for an isolated 2-circle system, the circles are twice as likely to be the same as they are to be different.

\begin{figure}[htb]
\centerline{\includegraphics[width=.5\textwidth]{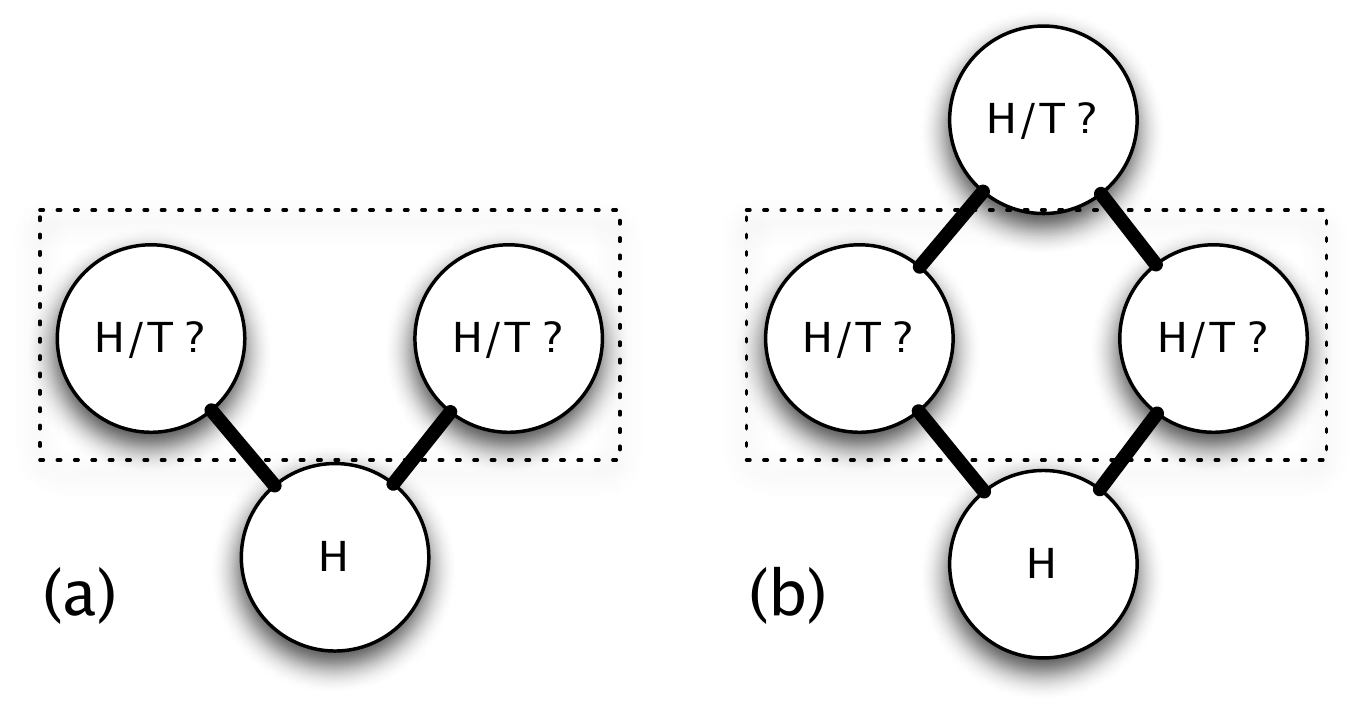}}
\end{figure}
 
In Figure 2a, given that the bottom circle is H, there are 4 different microstates compatible with an H on the left and an H on the right.  This is because there are two links, and they can each be either blue or green. (Specifically, listing the states of the three circles and the two links, the 4 possible ``HH" microstates are HBHBH, HBHGH, HGHBH, and HGHGH.)  According to the fundamental postulate of statistical mechanics, an HH will be four times as likely as a TT, for which only red links are possible (TRHRT).  The full table for Figure 2a is:
 
\begin{tabular}{cc|c}
\\
Left  & Right  & Microstates \\
\hline
H & H & 4 \\
H & T & 2 \\
T & H & 2 \\
T & T & 1 \\
\hline
2a & Total: & 9 \\
\\
\end{tabular}
 
Figure 2b is more complex, in that there is now a fourth circle at the top.  The fact that there are 4 links also means that there are 16 different microstates corresponding to all H's (4 green or blue links, $2^4=16$), but only one microstate corresponding to the case with T's on the right and left and another H on the top (4 red links).  The 2b table is:

\begin{tabular}{ccc|c}
\\
Left  & Right  & Top & Microstates \\
\hline
H & H & H & 16 \\
H & H & T & 4 \\
T & H & H & 4 \\
T & H & T & 4 \\
H & T & H & 4 \\
H & T & T & 4 \\
T & T & H & 1 \\
T & T & T & 4 \\
\hline
& 2b & Total: & 41 \\
\\
\end{tabular}

However, since we are not interested in the status of the top circle in this model, the relevant numbers are the total number of ways in which one might have (say) an H on the left and right.  To get the total number of such states, one simply sums the first two rows of the previous table.  In other words, there are 20 different states that have HH in the dotted box of Figure 2b; 16 with H on top and 4 with T on top.  The more useful 2b table is therefore:  

\begin{tabular}{cc|c}
\\
Left  & Right  & Microstates \\
\hline
H & H & 20 \\
H & T & 8 \\
T & H & 8 \\
T & T & 5 \\
\hline
2b & Total: & 41 \\
\\
\end{tabular}

Notice there are 25 ways in which the right and left circles match, vs. 16 ways in which they do not match.  This contrasts with a 5:4 ratio for Figure 2a.


\newpage

\begin{thebibliography}{}

\setlength{\baselineskip}{1.4\baselineskip} 

\section*{References}

\bibitem{Timpson}
C.G. Timpson, {\em Quantum Information Theory \& the Foundations of Quantum Mechanics}, Oxford (2013).

\bibitem{Wheeler}
J.A. Wheeler, ``Information, Physics, Quantum: The Search For Links" in {\em Complexity, Entropy and the Physics of Information},  W. H. Zurek (ed.), Addison-Wesley (1990).

\bibitem{Bell}
J. S. Bell, ``On the Einstein Podolsky Rosen Paradox", Physics {\bf 1}, 195 (1964).

\bibitem{KS}
S. Kochen and E.P. Specker, ``The problem of hidden variables in quantum mechanics", J. Math. Mech. {\bf 17}, 59 (1967).

\bibitem{PBR}
M.F. Pusey, J. Barrett, and T. Rudolph, ``On the Reality of the Quantum State'', Nature Physics {\bf 8}, 475 (2012).

\bibitem{FQXi4}
K. Wharton, ``The Universe is Not a Computer'', arXiv:1211.7081 (2012).

\bibitem{dBB}
D. Bohm, ``A Suggested Interpretation of the Quantum Theory in Terms of Hidden Variables", Phys. Rev. {\bf 85}, 166 (1952).

\bibitem{Everett}
H. Everett, ``Relative State Formulation of Quantum Mechanics'', Rev. Mod. Phys. {\bf 29}, 454 (1957).

\bibitem{Wharton13}
K.B. Wharton, ``Lagrangian-Only Quantum Theory", arXiv:1301.7012 (2013).

\bibitem{EPW}
P.W. Evans, H. Price, and K.B. Wharton, ``New Slant on the EPR-Bell Experiment", Brit. J. Found. Sci. {\bf 64}, 297 (2013).

\bibitem{Feynman}
R.P. Feynman, {\em The Character of Physical Law}, BBC Publications (1965).

\bibitem{Shimony}
A. Shimony, {\em Sixty-Two Years of Uncertainty: Historical, Philosophical, and
Physical Inquiries into the Foundations of Quantum Mechanics,} Plenum, New York,
(1990).

\bibitem{WMP}
K.B. Wharton, D.J. Miller, and H. Price, ``Action Duality: A Constructive Principle for Quantum Foundations", Symmetry {\bf 3}, 524 (2011).






\end{thebibliography}
\end{document}